%% file: eesop.tex
\documentclass[format=acmsmall, review=false, screen=true]{acmart}

\usepackage{booktabs} 

\usepackage{ragged2e}
\usepackage{array}
\usepackage{url}
\usepackage{subfig}

\newcolumntype{P}[1]{>{\centering\arraybackslash}p{#1}}

\newcommand\T{\rule{0pt}{2.6ex}}       
\newcommand\B{\rule[-1.2ex]{0pt}{0pt}} 

\begin{document}
\title[Scale-Out Processors \& Energy Efficiency]{Scale-Out Processors \& Energy Efficiency}  
\author{\small Pouya~Esmaili-Dokht}
\orcid{0000-0001-8799-5773}
\affiliation{%
  \institution{\scriptsize Universitat Polit\`ecnica de Catalunya (UPC) \& Barcelona Supercomputing Center (BSC)}
  \department{\scriptsize Department of Computer Science}
  \streetaddress{}
  \city{}
  \state{}
  \postcode{}
  \country{}}
\author{\small Mohammad~Bakhshalipour}
\affiliation{%
  \institution{\scriptsize Sharif University of Technology \& Institute for Research in Fundamental Sciences (IPM)}
  \department{\scriptsize Department of Computer Engineering \& School of Computer Science}
  \city{Tehran}
  \state{Tehran}
  \postcode{13416-45871}
  \country{Iran}}
\author{\small Behnam~Khodabandeloo}
\affiliation{%
  \institution{\scriptsize Institute for Research in Fundamental Sciences (IPM)}
  \department{\scriptsize School of Computer Science}
  \city{Tehran}
  \state{Tehran}
  \postcode{19538-33511}
  \country{Iran}}
\author{\small Pejman~Lotfi-Kamran}
\affiliation{%
  \institution{\scriptsize Institute for Research in Fundamental Sciences (IPM)}
  \department{\scriptsize School of Computer Science}
  \city{Tehran}
  \state{Tehran}
  \postcode{19538-33511}
  \country{IRAN}}
\author{\small Hamid~Sarbazi-Azad}
\affiliation{%
  \institution{\scriptsize Sharif University of Technology \& Institute for Research in Fundamental Sciences (IPM)}
  \department{\scriptsize Department of Computer Engineering \& School of Computer Science}
  \city{Tehran}
  \state{Tehran}
  \postcode{19538-33511}
  \country{IRAN}
}

\begin{abstract}
\input{abs}

\end{abstract}

\keywords{Scale-Out Workload, Data Center, Server Processor,  Performance per Watt, Performance per Unit Area.}

\maketitle

\renewcommand{\shortauthors}{}

\input{intro}

\input{method}

\input{eval}

\input{related}

\input{conc}

\medskip
\bibliographystyle{ACM-Reference-Format}
\bibliography{ref}

\end{document}

%% file: abs.tex
Scale-out workloads like media streaming or Web search serve millions of users
and operate on a massive amount of data, and hence, require enormous
computational power. As the number of users is increasing and the size of data
is expanding, even more computational power is necessary for powering up such
workloads. Data centers with thousands of servers are providing the
computational power necessary for executing scale-out workloads. As operating
data centers requires enormous capital outlay, it is important to optimize them
to execute scale-out workloads efficiently. Server processors contribute
significantly to the data center capital outlay, and hence, are a prime
candidate for optimizations. While data centers are constrained with power, and
power consumption is one of the major components contributing to the total cost
of ownership (TCO), a recently-introduced scale-out design methodology
optimizes server processors for data centers using performance per unit area.
In this work, we use a more relevant performance-per-power metric as the
optimization criterion for optimizing server processors and reevaluate the
scale-out design methodology. Interestingly, we show that a scale-out processor
that delivers the maximum performance per unit area, also delivers the highest
performance per unit power.

%% file: intro.tex
\section{Introduction}

Companies like Google, Facebook, and Microsoft rely on their data centers to
deliver scale-out services like streaming, social networking, and search. Such
high-throughput data centers consume enormous energy while executing scale-out
applications. As such, data centers consume more than three percent of total
global energy and contribute to two percent of the total CO2
emissions~\cite{IEEE:databaseenergy}. Economical and environmental
concerns necessitate making data centers more energy efficient.

Server processors contribute significantly to the power consumption of data
centers~\cite{barroso:datacenter:2nd}. Data centers use conventional server
processors~\cite{ferdman:clearing,reddi:web} where highly speculative
general-purpose processor cores are surrendered with large last-level caches
(LLCs). As technology scaling provides more transistors, more cores with highly
capable memory controllers and larger caches are employed in conventional
processors. Prior work reveals that cores' capabilities, off-chip memory
bandwidth, interconnection networks' bandwidth and cache size are over-provisioned
with respect to what scale-out workloads
need~\cite{ferdman:clearing,ferdman:mismatch,ferdman:case, bakhshalipour2018fast, bakhshalipour2018domino}. Accordingly, using
the conventional methodology for scale-out data centers is a poor choice with
respect to both performance and energy efficiency.

In another approach, tiled processors, which have many small and
energy-efficient cores, replace conventional processors for the purpose of
increasing per-server throughput as a result of using more processor
cores~\cite{mpr:tilera}. Just like conventional processors, more cores and
larger caches are placed in tiled processors as a result of technology scaling.
Although tiled processors improve energy efficiency and performance as compared
to conventional processors~\cite{lim:understanding}, they make suboptimal use
of silicon real estate~\cite{lotfi:sop}. Large caches found in tiled designs
are not effective for scale-out workloads because such caches are much smaller
than the size of the data sets and much larger than the instruction footprint.
So they cannot capture the data sets anyway and are beyond what is needed for
instructions. Not only large caches have long access latencies, but also they
impose high power usage. Moreover, in tiled processors, as the number of tiles
increases, the access latency of LLC also increases~\cite{mpr:tilera}.
Consequently, tiled methodology is not a good candidate for today's and
especially tomorrow's energy-efficient designs.

Recent work proposed a scalable processor architecture that is based on the
scale-out design methodology to maximize performance density (i.e., performance
per unit area)~\cite{lotfi:sop}. The building block of the resulting processors
named scale-out processors is a pod. A pod is a module that combines a few
cores with a small LLC to form a server. A pod runs an operating system and has
its own software stack. A scale-out processor consists of one or more pods with
no inter-pod connectivity. With scale-out processors, technology scaling
results in increasing the number pods. Prior work showed that scale-out
processors maximize performance density (PD)~\cite{lotfi:sop} and improve total
cost of ownership (TCO)~\cite{grot:tco} as compared to tiled and conventional
processors. 

Previous work optimized scale-out processors using performance per unit area
due to the importance of die area at 40~nm fabrication technology. But in
technologies below 20~nm, both at the chip level and at data centers, power and
energy are number one
constraints~\cite{esmaeilzadeh:dark,barroso:datacenter:2nd}. While scale-out
processors offer the highest performance density~\cite{lotfi:sop}, it is not
clear if they are optimal with respect to energy efficiency. To shed light on
this issue, in this work, we use a similar methodology as prior
work~\cite{lotfi:sop} but use \textit{performance per power $(P^3)$} as the
optimization criterion. 

Our experiments show that scale-out processors that are optimized for
performance density are also optimal with respect to energy efficiency and vice
versa. In this work, we make the following contributions:

\begin{itemize}

\item We use a system that consists of both processors and DRAM to evaluate the
energy efficiency of various processor organizations.

\item We show that for the technology node that we considered, the optimal pod
configuration using performance per power is the same as what has been obtained
using performance density.

\item We show that the optimal pod configuration does not change for a large
variety of technology nodes and DRAM parameters.

\end{itemize}

%% file: method.tex
\section{Methodology}
\label{method}

Prior work~\cite{lotfi:sop} came up with a methodology to allocate the limited
resources (mainly area) to the various components of a multi-core processor
targeting maximization of throughput per unit area. In this work, we attempt to
efficiently allocate power to various components targeting maximization of
energy efficiency. Furthermore, we discuss how the optimal pod configuration
changes if various characteristics of the system change. We use a combination
of cycle-accurate simulation, analytic models and technical reports for this
study.

\subsection{Design and technology parameter}

We analyze various designs in 14~nm technology using 0.8 volts for chip supply
voltage. Our area constraints set to 280~$mm^2$, our power budget for all
designs sets to 95~W. We also use up to 6 single-channel DDR4 interfaces in
our chip designs.

\begin{table}[!t]
\caption{Estimated power and area values for different components at 14nm}
\label{table_example}
\centering
\begin{tabular}{| c | c | c | c |}
\hline
\multicolumn{2}{|c|}{Components} & Area & Power\T\B\\
\hline
Cores&\begin{tabular}{@{}l@{}}
Conventional\T\\
Out-of-Order\\
In-Order\B
\end{tabular}& \begin{tabular}{@{}l@{}}
3.1~$mm^2$\T\\
1.1~$mm^2$\\
0.32~$mm^2$\B
\end{tabular}&\begin{tabular}{@{}l@{}}
3.8~W\T\\
0.4~W\\
0.2~W\B
\end{tabular}\T\\
\hline
LLC & 16-way SA & 0.62~$mm^2$ Per MB & 0.2~W per MB\T\B\\
\hline
\multicolumn{2}{|c|}{Interconnect} & 0.2-4.5~$mm^2$ & $<$5~W\T\B\\
\hline
\multicolumn{2}{|P{3cm}|}{
\begin{tabular}{@{}l@{}}
DDR4 controller\T\\
(PHY+controller)\B
\end{tabular}
} & 12~$mm^2$ & 5.7~W\T\B\\
\hline
\multicolumn{2}{|c|}{SoC components} & 42~$mm^2$ & 5~W\T\B\\
\hline
\end{tabular}
\label{tab:estimatedparameter}
\end{table}

Table~\ref{tab:estimatedparameter} contains a summary of design parameters.
Reported powers are estimation of real power on our workloads. We use three
different core types in our study. Conventional processors represent
aggressive, 4-way, highly speculative core microarchitecture. Tiled and
scale-out processors are evaluated by two different core types. The first model is a
3-way high-performance out-of-order core representing ARM
Cortex-A15~\cite{mpr:a15} and the second model is a dual-issue in-order core,
similar to Cortex-A8~\cite{mpr:a8}. We set all cores' frequency to 2~GHz in
order to make comparison between different architectures convenient. Cache area
and power parameters are extracted from CACTI 6.5~\cite{muralimanohar:cacti}.

Area of different SoC components derived from scaling micrograph of a Nehalem
processor in 45~nm technology~\cite{kumar:nehalem}. We extract DDR4 DRAM power
consumption parameters using published DDR4 power
characteristics~\cite{dram:ddr4}. As DRAM cannot easily be scaled beyond
20~nm technology~\cite{dramscale:power,dramscale:power2}, we assume 20~nm DRAM in this study.
 We consider at most 70\% utilization for DDR4 memory
channels~\cite{david:memory}. Power of other SoC components estimated by
modeling Sun UltraSPARC T2 configured using McPAT v0.8~\cite{li:mcpat}. 

\subsection{Chip organizations}

For each design, we use as many cores and as much cache as we can without violating any
constraints in area, power or memory bandwidth. Maximum required memory
bandwidth determines the number of memory controllers in our designs.
Performance and power estimation methodologies are described in
Sections~\ref{perf} and~\ref{power}, respectively.

\subsubsection{Conventional}

Conventional processor can accommodate at most 17 cores before reaching the
specified power budget. Three DDR4 memory controllers are sufficient to serve
the off-chip demands. We use 48~MB of LLC in the processor. Cores and caches
are connected through a crossbar interconnect.

\subsubsection{Tiled with OoO cores}

Tiled OoO processor can accommodate 139 cores before reaching the power
constraint. We use 80~MB of LLC in this processor. A mesh interconnect with
3-cycle delay per hop for both link and router is used for all tiled designs.

\subsubsection{Tiled with in-order cores}

By keeping the same LLC size as tiled OoO design, tiled
in-order design has 225 cores and 80~MB of LLC. In this design, power
constraint restricts the number of cores.

\subsubsection{Scale-out}

For determining core count and cache size of scale-out design, we have done an
extensive evaluation changing the cache from 1 to 8~MB and core count from 1 to
256.

\subsection{Scale-out workloads}

We take scale-out workloads from CloudSuite. Our workloads include Data
Serving, MapReduce, SAT Solver, Web Frontend, and Web Search. We have two
MapReduce workloads in our suite, classification (MapReduce-C) and word count
(MapReduce-W).

\subsection{Performance evaluation}
\label{perf}

As cycle-accurate full-system simulation is 100,000 times slower than real
hardware~\cite{wenisch:simflex}, it is impractical to search the whole design
space with time-consuming simulations. Instead, we use an analytic
model~\cite{hardavellas:phdthesis,lotfi:sop} that its parameters derived from
simulations. This model predicts performance based on cache size, cache miss
ratio, core count, cache access latency and memory access latency. 

To derive parameters of the analytic model, we use full-system simulation. For
full-system simulation of different pod sizes, we use
Flexus~\cite{parsa:flexus}, which is built on top of Virtutech Simics. Flexus
extends Simics functional model with detailed models of OoO and in-order cores
and the cache hierarchy. 

We evaluate 10 seconds of execution of each workload using SimFlex sampling
methodology~\cite{wenisch:simflex}. For each measurement, we load checkpoints
with warmed caches and branch predictors, and then run 100~K cycles to reach
the steady state before collecting measurements for the subsequent 50~K cycles.
For Data Serving workload, we need to run the simulations for 2000~K cycles to
reach the steady state. We use the ratio of the number of application
instructions to the total number of cycles (including the cycles spent
executing operating system code) to measure performance; this metric has been
shown to accurately reflect overall system throughput~\cite{wenisch:simflex}.
Throughput measured with 95\% accuracy and an average error rate lower than
4\%.

\subsection{Power evaluation}
\label{power}

We use McPAT for power estimation of SoC components. For cores, however, recent
studies show that McPAT is not accurate for power analysis due to the
differences between core structure and its implementation~\cite{xi:mcpaterror}.
As an alternative, prior work has shown that Instruction per Cycle (IPC) is
strongly correlated to the power consumption~\cite{rodrigues:corepower,
contreras:power, bircher:complete}. For example, Bircher and
John~\cite{bircher:complete} report an average of only 3\% error in core's
power usage when compared to the measured CPU power. Moreover, Rodrigues et al.
show that it is possible to estimate a core's power usage with an average error
rate of less than 3.9\% using performance counters~\cite{rodrigues:corepower}.
Using these approaches requires having power and energy numbers of the examined
cores. For this purpose, we use the empirical power reports from the published
technical report~\cite{vasilakis:armpower} for cores' power estimation.

%% file: eval.tex
\section{Results}

We first find the optimal pod size for each core type and then replicate pods
in each design to reach one of the constraints. Subsequently, we compare the resulting
scale-out processor with tiled and conventional architectures. Finally, we
compare performance-density optimal scale-out processors against their
performance-per-power optimal counterparts.

\begin{figure*}[t]
\centering
\includegraphics[width=1\textwidth]{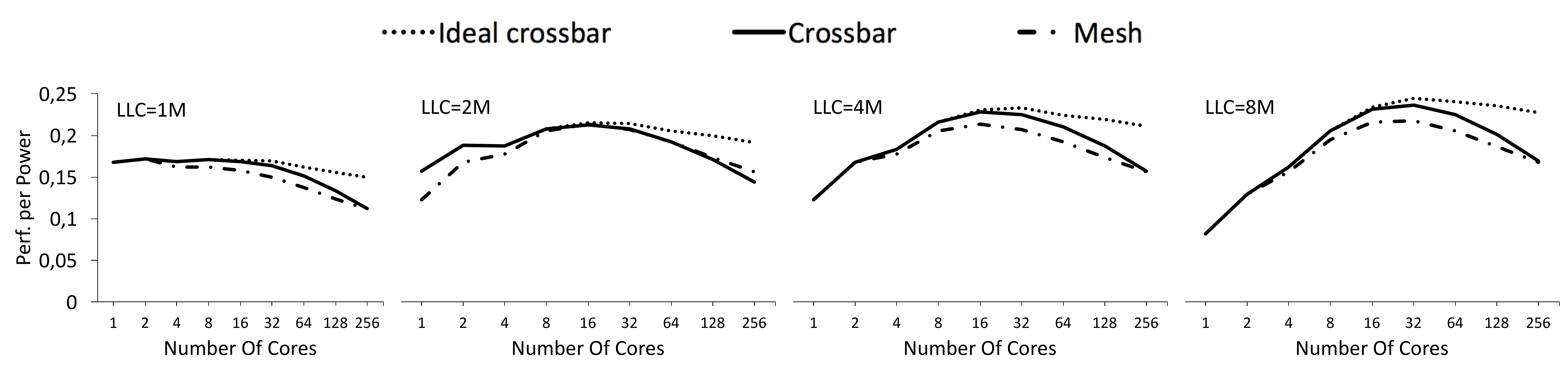}
\caption{Performance per power ($P^3$) for a design with OoO cores and various cache sizes}
\label{fig:OoOcoreheader}
\end{figure*}

\subsection{System with out-of-order cores}

Average performance per power ($P^3$) across all workloads for four different
LLC sizes is shown in Figure~\ref{fig:OoOcoreheader}. Larger cache sizes are
not investigated as they deteriorate performance per power.  Each graph
contains three lines corresponding to three different interconnect types.

We observe that in all designs and regardless of cache size and interconnect,
performance per power diminishes as the number of cores starts to exceed 32.  A
system with 16 cores, 4~MB of LLC and a crossbar interconnect maximizes $P^3$.
This is identical to the pod that maximizes performance per unit
area~\cite{lotfi:sop}. 

Based on circumstances discussed in Section~\ref{method}, our scale-out
processor design at 14~nm is power-limited and can accommodate eight pods. The
resulting system area and power are 253~$mm^2$ and 87~W (with DRAM 130~W),
respectively.

Scale-out design with out-of-order cores achieves $3.95\times$ higher $P^3$ as
compared to the conventional processor due to using simpler cores and a smaller
LLC. Also, a scale-out design has notable advantages over tiled designs with
respect to $P^3$: its overall $P^3$ is 26\% higher than the tiled design. This
advantage stems from inefficient large cache size and long inter-hop latency in
tiled architectures.

\begin{figure*}[!t]
\centering
\centerline{\includegraphics[width=6in]{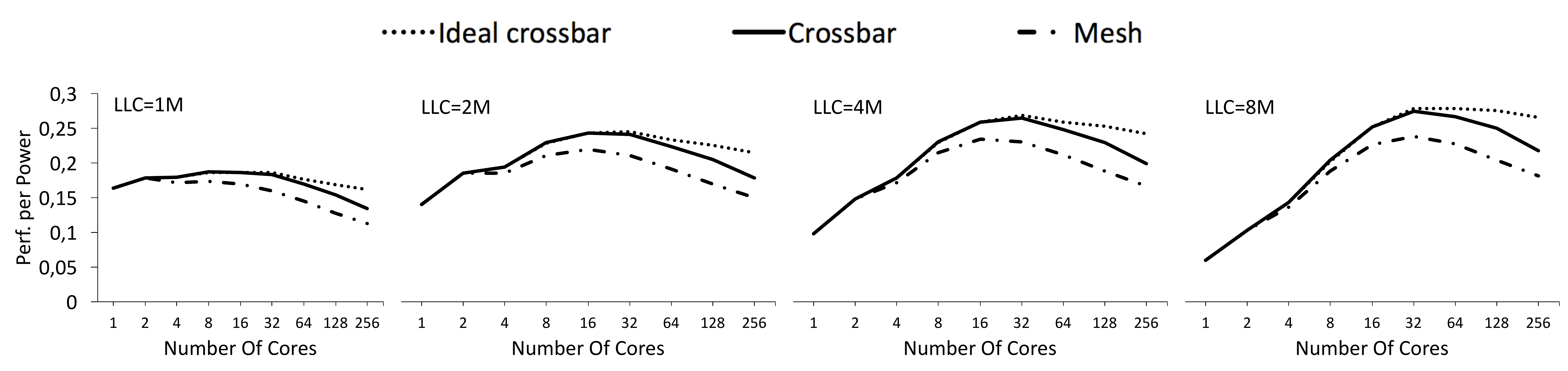}}
\caption{Performance per power ($P^3$) for a design with in-order cores and different cache sizes}
\label{fig:In-Ordercoreheader}
\end{figure*}

\subsection{System with in-order cores}

Figure \ref{fig:In-Ordercoreheader} shows the average performance per power of
different processors across all workloads. Based on these results, a
$P^{3}$-optimal pod contains 32 cores with 4~MB of LLC and a crossbar
interconnect. Again, the $P^{3}$-optimal pod with in-order cores is identical
to the performance-density optimal pod~\cite{lotfi:sop}. This is because
scale-out processors are tuned for the characteristics of scale-out workloads:
(1) massive request-level parallelism, (2) large instruction footprint, and (3)
enormous datasets in the main memory.

Resulting scale-out processor with in-order cores can afford seven pods before
violating the power budget. With all peripherals and interconnect, scale-out
chip's total die-area is 193~$mm^2$ and consumes 86~W (with DRAM 139~W).

The scale-out chip with in-order cores offers 43\% higher $P^3$ as compared to a
tiled design. Furthermore, it achieves $3.2\times$ higher $P^3$ over conventional
designs.

\begin{figure}[!t]
\centering
\subfloat[]{\includegraphics[width=3in]{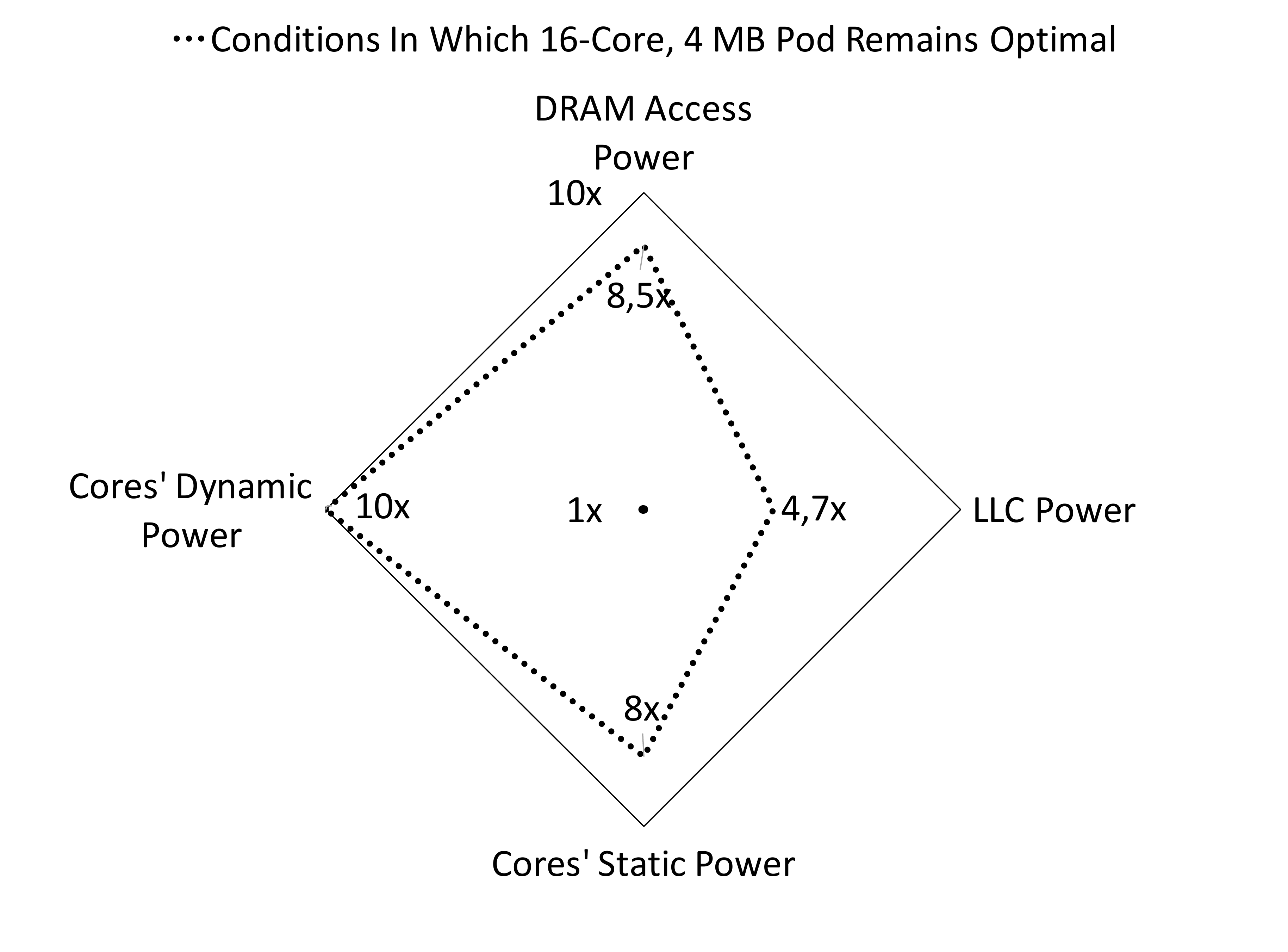}%
\label{fig:IncPodSize}}

\subfloat[]{\includegraphics[width=3in]{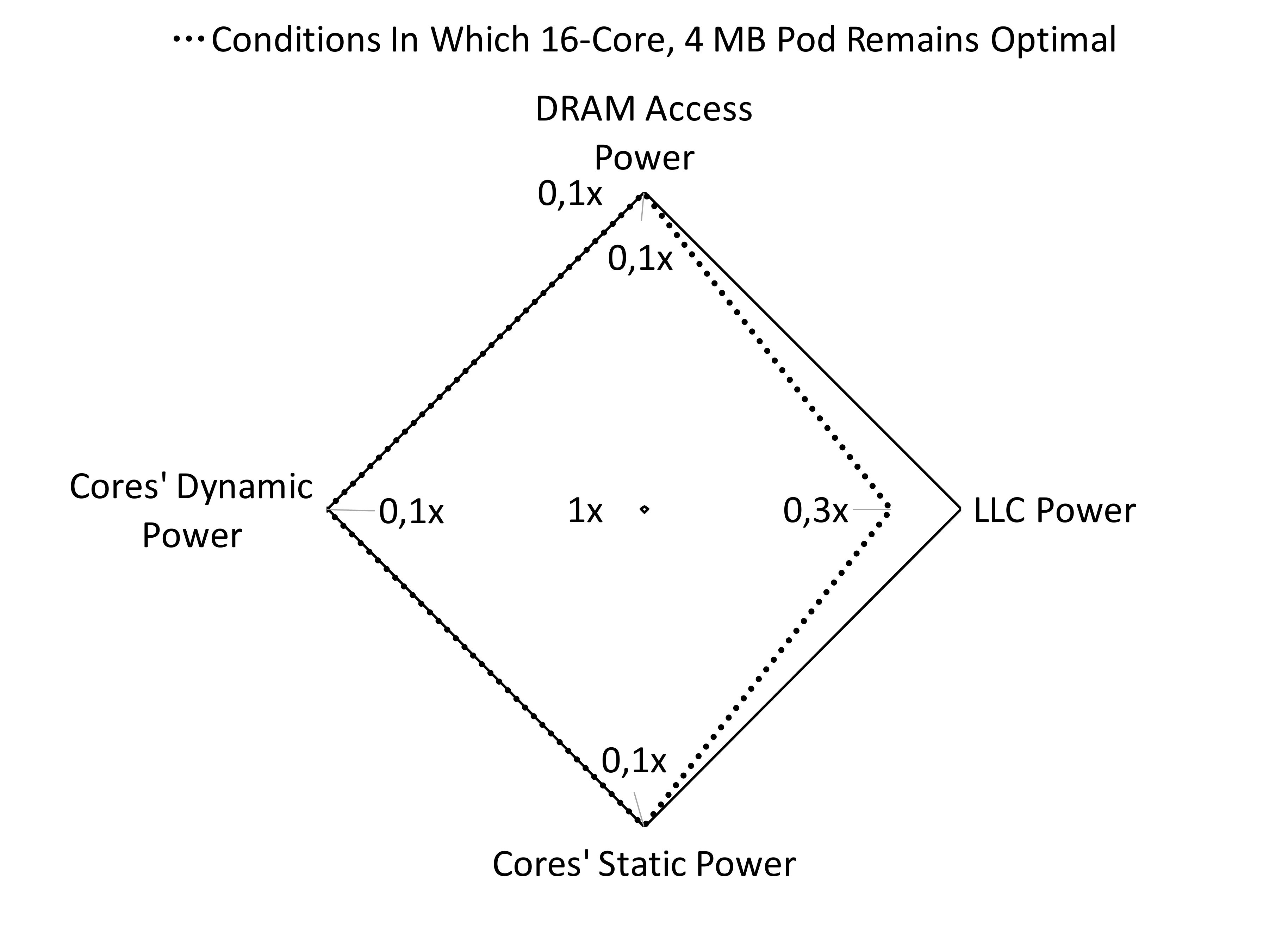}%
\label{fig:DecPodSize}}
\caption{We vary cores' dynamic power, cores' static power, LLC power usage, and DRAM access power by up to 10$\times$ up (down) with respect to their current values for a scale-out processor designed with OoO cores in Part a (b). The solid rectangles show the state space. The dotted rectangles show parts of the state space in which the 16-core, 4-MB pod remains optimal.}
\label{fig_Podsize}
\end{figure}

\subsection{Sensitivity of Optimal Pod Configuration to Design Parameters}

We perform a study on how the optimal pod will change if parameters of the
design change. We use OoO small cores in this study. All the remaining design
aspects are the same as previous experiments. LLC power usage, core's static
and dynamic power, and DRAM access energy are the main elements of this study.
We sweep the energy usage of these components from $0.1\times$ to
$10\times$ of the current values to see how these changes affect the optimal
pod configuration.  Figure~\ref{fig_Podsize} shows the results of our study.
The solid rectangles indicate the state space while the dotted rectangles show
parts of the state space in which the optimal pod configuration does not
change. The figure clearly shows that the 16-core, 4-MB pod remains the optimal
pod configuration for a large range of parameters. 

Figure~\ref{fig:IncPodSize}~~shows that changing cores' dynamic power by
10$\times$ does not change the optimal pod configuration. Moreover, cores'
static power affects the optimal pod configuration only when it is increased by
$8\times$ of its current value. Power-hungry cache system that at least
consumes $4.7\times$ more power, moves us towards having a smaller pod with
fewer cores and a smaller LLC. On the other hand, increasing the DRAM access
energy by more than $8.5\times$ does the exact opposite. A power-hungry DRAM
calls for a pod with a larger LLC to filter out more data accesses.

Figure~\ref{fig:DecPodSize}~~shows that a 10$\times$ decrease in core power or
DRAM access energy does not change the optimal pod configuration. Moreover, a
low-power LLC only affects the optimal pod configuration when its power usage
becomes $0.3\times$ of its current value. This means that in more advanced
technology nodes in which the energy of the core, cache, and DRAM is not expected
to change significantly, the optimal pod configuration is likely to remain the same. 

\begin{table*}[!t]
\caption{Resulting chips at 14~nm technology}
\label{table_example}
\centering
\resizebox{\linewidth}{!}{%
\begin{tabular}{|l|c|c|c|c|c|c|c|c|c|}
\hline
\multicolumn{10}{|c|}{14~nm}\T\B\\
\hline
Processor Design     & Constraint    & Cores & LLC (MB) & MCs & Area ($mm^2$) & Performance & Power (Watt) & PD   & $P^3$\T\B\\
\hline
Conventional         & Power-limited & 17    & 48       & 3   & 161         &23        &105        & 0.14 & 0.22\T\B\\
\hline
Tiled (OoO)          & Power-limited & 139   & 80       & 3   & 280        &86&128&0.31&0.67\T\B\\
\hline
Scale-Out (OoO)      & Power-limited & 128   & 32       & 5   & 253        &109&130&0.43&0.84\T\B\\
\hline
Tiled (In-Order)     & Power-limited & 225   & 80       & 5   & 224        &80&137&0.36&0.58\T\B\\
\hline
Scale-Out (In-Order) & Power-limited & 224   & 28       & 6   & 193        &116&139&0.60&0.83\T\B\\
\hline
\end{tabular}}
\label{tab:14nm}
\end{table*}

\subsection{Summary}

Table$~$\ref{tab:14nm} summarizes our chip-organization, power consumption,
limiting factor, area, performance, power, PD and $P^3$ in 14~nm technology.
As we consider DRAM dynamic power in our study, reported powers are more than
the power budget that we set in Section~\ref{method}, however, all chips
consume less power than the limit. Performance column shows average
user-instruction per clock cycle~\cite{wenisch:simflex} that the corresponding
design can deliver.

Our study indicates that a single pod configuration is optimal for both energy
efficiency and performance density. Also, many technological changes in the
cache, core or DRAM do not change the optimal pod configuration. We also
showed how the pod configuration would change if characteristics of the
components change significantly.

%% file: related.tex
\section{Related Work}

There are proposals that optimize data-center cost, power, and/or area with an
efficient processor architecture. Such pieces of prior
work~\cite{kgil:picoserver, hardavellas:toward, lotfi:sop, grot:tco} partially
share some of the insights and/or conclusions of this work. Our work is
different from prior work on scale-out processors~\cite{lotfi:sop,grot:tco} in
many aspects. Unlike those studies that target area as the optimization
criterion, we use energy efficiency. While prior work~\cite{grot:tco} showed
that a performance-density (PD) optimal processor also offers better energy
efficiency, this work is the first to show that a PD-optimal processor is
also optimal with respect to energy efficiency.  Moreover, unlike prior work,
we included DRAM energy in our study.  Finally, we study the effect of
variations of subsystem characteristics on the optimal pod configuration.

%% file: conc.tex
\section{Conclusion}

As the primary constraint of data centers is power usage, server processors that
are optimized for scale-out workloads should exhibit excellent energy efficiency.
For this purpose, we revisited the scale-out design methodology with respect to
energy efficiency. We found that in many real-world conditions (like the ones in
our study), the scale-out processors that are optimal with respect to
performance density are also optimal with respect to energy efficiency.